\documentclass[10pt,paper,english]{article}
\usepackage{babel}
\usepackage[latin1]{inputenc}
\usepackage{graphicx}
\usepackage{amscd}
\usepackage{graphicx}
\usepackage{amsmath,amsthm}
\usepackage{graphicx}
\usepackage{epstopdf}
\usepackage[all]{xy}
\usepackage{gastex}
\usepackage{color,graphicx}
\usepackage{graphics}
\usepackage{amsmath,amssymb,amsthm}
\usepackage{makeidx}
\usepackage{enumerate}
\usepackage{setspace}
\usepackage{array}
\usepackage{float}
\begin{document}

\theoremstyle{definition}
\newtheorem{thm}{Theorem}
\newtheorem{lem}[thm]{Lema}
\newtheorem{definition}[thm]{Definition}
\newtheorem{remark}[thm]{Remark }
\newtheorem{teo}[thm]{Theorem}
\newtheorem{exa}[thm]{Example}
\newtheorem{co}[thm]{Corollary}
\newtheorem{prop}[thm]{Proposition}
\newtheorem{ta}[thm]{Table}

\date{}
\title{Indirect Influences in International Trade}
\author{Rafael D\'iaz \ \ \  and \ \ \   Laura G\'omez}
\maketitle
\begin{abstract}
We address the problem of gauging the influence exerted by a given country on the global trade market from the viewpoint of complex networks. In particular, we apply the PWP method for computing indirect influences on the world trade network.
\end{abstract}

\section{Introduction}

Global and local, private and public institutions increasingly rely on numerical indexes for their decision making processes.
A single index may be the difference between approval or rejection. These indexes, and even the mathematical models behind them, are often made public to enhance transparency, accountability, and establish standards. The  usefulness of an index
is measured by the successes and failures of decisions based on it. Thus index creation, computation, selection, comparison,
evaluation, and consolidation has become an industry of its own, and an integral part of institutionalized decision making processes.\\

In this work we address the problem of finding suitable indexes for ranking countries according to their influences on the
international trade market. In this case there is an obvious candidate: the influence of a country on the international trade market is proportional to the amount (counted in US dollars) of its international trade, i.e. the total amount of imports plus the total amount of exports. This primordial index is sound and should not be overlooked. Nevertheless, we claim that it disregards some subtle but important issues.\\

Suppose we have a couple of countries $A$ and $B$ both with high levels of international trade, so that they both are
highly ranked with the above index. Suppose in addition that $A$ and $B$ trade essentially with each other, i.e. trade of $A$
and $B$ with other countries is negligible in comparison with trade between themselves. In this case our feeling is that
$A$ and $B$ do not exert a strong influence on the international trade market: a disruption on $A$'s economy will surely impact $B$'s economy, but will have a negligible impact on global trade. Countries $A$ and $B$ although highly interdependent are in fact quite isolated from the rest of the world, and should not be ranked as highly influential countries in the international trade market.\\

To address this sort of issues we take a network approach \cite{f2, verde} to our index creation problem. In Section \ref{stn}, we regard the international trade market as a weighted network with nodes representing countries, edges representing trade between countries, and weights measuring the influence that a country exerts on another country trough trade. Indeed, in Sections \ref{dit} and \ref{dio}, we are going to introduce a couple of different weights leading to a couple of rankings.\\

Once we have established the network settings for approaching our index creation problem,  we face the
problem of ranking nodes in a network by their influence.  This demands that we paid attention to the distinction between direct and indirect influences in a network, a distinction emphasized  by Godet and his collaborators \cite{g1},
who stressed  the  power of studying indirect influences for uncovering hidden relations. Direct influences in a network
arise from directed edges. Indirect influences in a network arise from chains of direct influences, i.e. from directed paths. \\

Let us consider an extreme case that illustrates the importance of taking indirect influences into account.
Cuba and the United States (US) have a fairly weak amount of trade, indeed since 1960  the US has placed
a series of economic, financial, and commercial  prohibitions to trade with Cuba. Thus a fairly low influence
of the US on the Cuban economy is to be expected. However, the US trades with Cuba's main
trading partners, and so we may expect that the US exerts a stronger indirect influence on the Cuban economy  than what
one may naively think. Simply put, a disruption of the US's economy will likely impact the Spaniard economy, and trough Spain
it will also be felt by the Cuban economy. The main challenge that we confront in this work is to give a quantitative account of the latter impact. \\

After building our mathematical models in Sections \ref{stn} trough \ref{hiitn}, we proceed in Sections \ref{secactn} and  \ref{wtn2} to implement them using real world data. Out of United Nations member
states we restrict our attention to 177 countries, namely, those that have recent data available in the Economic Commission for Latin America and the Caribbean (ECLAC) web site \cite{pag}.  We collect the  2011 data of exports and imports between each pair of countries, as well as the GDP, and the exports and imports totals for each country.  Using this data we build the "international trade network," where nodes are countries, and a pair of nodes is connected
by an edge if there were exports or imports between them in 2011.  Suppose there actually was trade between countries $A$ and $B$,
then we define a couple of weights on the edge from $A$ to $B$ as follows:

\begin{enumerate}
\item \textbf{Direct Influences on Trade.} This weight computes the proportion of the international trade of $B$ (exports + imports) that involves country $A.$\\

\item \textbf{Direct Influences on Offer.} This weight computes the relative contribution of the trade between $A$ and $B$ to
the offer of $B$ (GDP + imports).
\end{enumerate}

To take indirect influences into account we start from a trade network, with one of the weights just introduced,
and apply one of the mathematical methods available for computing indirect influences in complex networks. With any of these methods one goes beyond computing direct trade between countries $A$ and $B$, and takes as well into account trade chains $$A=A_1,\ A_2,\ ....\ ,A_n=B$$ such that country $A_i$ trades with country $A_{i+1}$.\\

In this work we weight chains of trade using the PWP method introduced by D\'iaz \cite{r1}. To place this method whiting a general context and for the reader convenience we provide a in Section \ref{mcii} a succinct description of four closely related methods for computing indirect influences, namely, Godet's MICMAC, Google's PageRank, Chung's Heat Kernel, and the PWP method. For more details on the similarities and differences among these methods the reader may consult \cite{r1}.  Our choice of the PWP method rest on the fact that with it any chain of direct influences, of any length, generates indirect influences, and reciprocally, all indirect influences are generated in this fashion. \\

Thus we obtain several rankings among nations using the methodology outlined above: for each of the above weights we get the direct and indirect influences rankings. We compare these rankings among themselves and with the GDP ranking for the American continent trade network in Section \ref{secactn},  and for the World trade network
in Section \ref{wtn2}. We show that there are remarkable  differences between the various rankings, and analyse them in economic terms.\\

We remark once again that our economic data refers to year 2011, and is measured in US dollars.
When we compare different indexes what is really at stake is to compare the rankings they induce. In practice what we do
is to compare the corresponding normalized indexes. Numerical calculations in this work were made with Scilab's module for computing indirect influences designed by Catumba \cite{pag2}. For the economics terminology the reader may consult
\cite{m12}.

\section{Trade Networks}\label{stn}

In this section we introduce basic definitions  concerning the construction of international
trade networks and their adjacency matrices, called matrices of direct influences in this work.

\begin{definition}
A trade network $G=(V,E)$ is a directed graph such that:
\begin{itemize}
\item The set $V$ of vertices is a family of countries.
\item There is an edge in $E$ from vertex $A$ to  vertex $B$  if and only if there is trade (exports or imports) between $A$ and $B$.
\end{itemize}
\end{definition}

Note that trade networks are actually symmetric graphs. Nevertheless, we regard them as directed graphs since we are soon going to introduced non-symmetric weights on them. For each edge $e \in E$, let $s(e)$ be its source vertex, or equivalently, the country that exerts the influence, and let $t(e)$ be its target vertex,  or equivalently, the country that is influenced. A
weight $w$ on a directed graph $G$ assigns weight $w(e) \in \mathbb{R}$ to each edge of $G$, i.e. a weight on $G$ is a map
\begin{center}
$w : E \ \longrightarrow \ \mathbb{R} $.
\end{center}
In Sections \ref{dit} and \ref{dio} we are going to introduce a couple of weights on trade networks.\\

The bi-degree $(i(A),o(A)) \in \mathbb{R}\times \mathbb{R}$ of a vertex $A \in V$ in a weighted directed graph $G$ is such that $i(A)$ is the sum of the weights of edges with target $A$, and  $o(A)$ is the sum of the weights of  edges with source $A$, that is:
$$i(A)\ = \ \sum_{e \in E, \ t(e)= A }  w(e) \ \ \ \ \ \ \ \ \mbox{and} \ \ \ \ \ \ \ \ o(A)\ = \
\sum_{e \in E, \ s(e)= A}w(e).$$

We impose the alphabetic order on countries, so the set of countries $V$
is identified with the set $\{1,...,n \}$, where $n$ is the number of countries in our trade network.

\begin{definition}
The adjacency matrix or matrix of direct influences $D$ of a trade network $G$ is given for $A, B \in V$ by:
$$D_{AB} \ = \ D(A,B) \ \ =  \ \  \sum_{s(e)= B, \ t(e)= A} w(e).$$
\end{definition}
Thus $D_{AB}$ gives the direct dependence of $A$ on $B$, or equivalently, the direct influence of $B$ on $A$.
The matrix of indirect influences $T=T(D)$ is computed from the matrix of direct influences $D$ using one
of the methods discussed in Section \ref{mcii}. In our applications, the matrix of indirect influences  $T$ is computed applying
the PWP method.

\begin{definition} \label{d1}
Let $G$ be a trade graph, $D$ its matrix of direct influences, and $T=T(D)$ its associated matrix of indirect influences.
The indirect dependence $d_{A}$, indirect influence $f_{A}$, and indirect connectedness $c_{A}$ of vertex $A$ in $G$ are given, respectively, by:
$$d_{A} \ = \ \sum_{B \in V} T_{AB},  \ \ \ \ \ \ \ f_{A}\ = \ \sum_{B \in V}T_{BA},
\ \ \ \ \ \ \ \mbox{and} \ \ \ \ \ \ \ c_{A}\ = \ d_{A}\ + \ f_{A}.$$
The ordered pair $(d_{A}, f_{A})$ is the bi-degree of $A$ in the network of indirect influences. The indirect connectedness of vertices
$A$ and $B$ is given by $c_{AB}=T_{AB} + T_{BA}$
\end{definition}

\begin{remark}
Direct dependencies, influences, and connectedness are computed in a similar way
using the matrix of direct influences $D$ instead of the matrix $T$ of indirect influences.
\end{remark}

\begin{figure}[h]
\begin{center}
\includegraphics[scale=0.4]{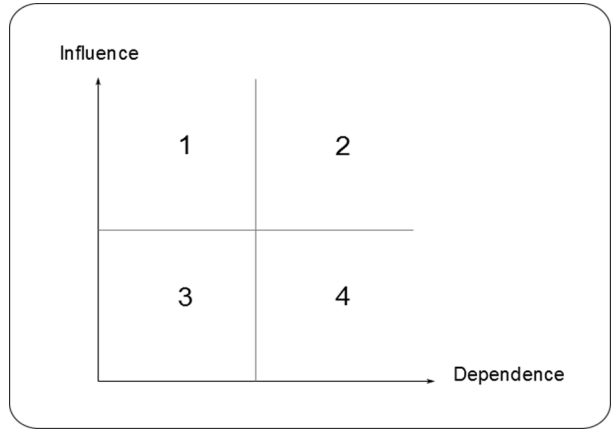}
\caption{Dependence-Influence Plane.}
\label{p1}
\end{center}
\end{figure}

Following Godet a $2$-dimensional representation of vertices bi-degrees can be displayed trough
the dependence-influence plane which comes naturally divided in four sectors, see Figure \ref{p1}.
The horizontal axis represents dependencies and the vertical axis represents influences. A country $A$ is represented
in the dependence-influence plane by the ordered pair $(d_{A}, f_{A})$. The horizontal and vertical lines defining the four sectors of the plane are located at the mean  dependence $\overline{d}$ and at the mean influence $\overline{f}$, given respectively, by
$$\overline{d} \ = \ \frac{1}{n}\sum_{A \in V}d_A \ \ \ \ \ \ \ \mbox{and} \ \ \ \ \ \ \
\overline{f} \ = \ \frac{1}{n}\sum_{A \in V}f_A  .$$

\begin{itemize}
\item \textbf{Sector 1:} Influential independent countries.

\item \textbf{Sector 2:} Influential dependent countries.

\item \textbf{Sector 3:}  Low influence independent countries.

\item \textbf{Sector 4:} Low influence dependent countries.

\end{itemize}

\section{Computing Indirect Influences}\label{mcii}

Although our focus is on the PWP method, for the reader convenience we introduce four methods for computing indirect influences, and describe how each method computes the matrix of indirect influences. For more on the similarities and differences among these methods the reader may consult \cite{r1}.

\subsection{MICMAC}

With this method, introduced in 1992 by Godet \cite{g1}, the matrix of indirect influences is  $T(D)=D^{k},$
where $D$ is the matrix of direct influences, and $k \in \mathbb{N}$ is a parameter usually equal to $4$ or $5$. The relevant paths in the network, with this method, are those of length $k$.

\subsection{PageRank}

This method, registered in 1999 by Google \cite{b2, b1, m1}, is quite well-known
thanks to its application to web searching.  With PageRank
influences are normalized, and dependencies measure the relevance or relative importance of web pages.
PageRank takes infinite powers of matrices, thus it gives greater importance to infinite long paths.
Let $D$ be the matrix of direct influences, whose entries should be non negative real numbers, and
the sum of each column must be either $0$ or $1$. The matrix of indirect influences is given by

\begin{center}
 $T(D) \ = \ \displaystyle \ \lim _{k \rightarrow \infty} [p \overline{D} +(1-p)E_{n}]^{k}$,

\end{center}
where $\overline{D}$ is the matrix obtained form $D$ by replacing the entries of a zero column by $\frac{1}{n}$,
 $E_{n}$ is the matrix with entries  $\frac{1}{n}$, and the parameter $p$ is usually $0.86$. \\

In the web network vertices are  web pages, and there is an edge from page $B$ to page $A$ if there is a hyperlink in $B$ leading to $A$. The matrix of direct influences $D$ is given by:
$$D_{AB}\ = \
\left\{ \begin{array}{lcl}
\frac{1}{o(B)} \ \   \mbox{ if  there  is  an  edge  from} \ B \ \mbox{to} \ A,\\
& & \\
0 \ \ \ \ \ \mbox{ otherwise},
\end{array}
\right.$$
where $o(B)$ is out-degree of the vertex $B$.\\

The  ranking of page $A$ is proportional to its indirect dependence $d_A$. A higher rank means an earlier showing in a web search. 	

\subsection{Heat Kernel }

This method was introduced in 2007 by Chung \cite{f1}. Let $D$ be the matrix of direct influences,
then the matrix of indirect influences is given by $$T(D) \ = \ e^{ \lambda(D-I)} \ = \  \sum _{k=0}^{\infty} (D-I)^{k}\frac{\lambda^k}{k!},$$
where $I$ is the identity matrix with the size of $D$, and $\lambda$ is a fixed positive real number.

\subsection{PWP}

This method  was introduced in 2009 by D\'iaz \cite{r1}. With PWP, as well as with MICMAC and  Heat Kernel, one can compute indirect influences for any matrix of direct influences, even one with negative entries.
Fix a positive real number $\lambda$ (set to $1$ in our applications) and let $D$ be the matrix of direct influences. The matrix of indirect
influences is given by: $$T(D) \ = \ \frac{e^{\lambda D}_{+}}{e^{\lambda}_{+}},$$ where for a  matrix or a number $x$ we set
$$ e^{x}_{+} \  =  \ e^{x}-1  \ = \  \sum _{k=1}^{\infty} \frac{x^{k}}{k!}.$$

\section{Direct Influences on Trade}\label{dit}

We go back to the problem of  defining direct influences on a trade network $G$.

\begin{definition}\label{ddit}
The direct influence on trade $C_{AB}$ of country $B$ on country $A$ is given by:
$$C_{AB}\ = \ \dfrac{E(B,A)\ + \ I(B,A) }{E(A)\ + \ I(A)}$$   where:
\begin{itemize}
\item $E(B,A)$ is the amount of exports from $B$ to $A$, according to $A$'s data.
\item $I(B,A)$ is the amount of imports from $B$ to $A$, according to $A$'s data.
\item $E(A)$ and $I(A)$ are, respectively, the total exports and total imports of  $A$.
\end{itemize}
\end{definition}

The direct influence on trade $C_{AB}$ measures the portion of $A$'s international trade that involves $B$: out of each dollar the $A$ trades with the world, it trades $C_{AB}$ dollars with $B$. A few remarks on the definition of  direct influences on trade $C_{AB}$:\\

\begin{itemize}
\item Exports from country $A$ to country $B$ are  often though as creating dependence of $B$ on $A$. In our approach
exports creates dependence in both directions: $B$ needs some products coming from $A$ (otherwise it would not buy), just
as $A$ needs the currency coming from $B$ (otherwise it would not sell.)\\

\item Dependencies are normalized, indeed we have that
$$d_A \ = \ \sum_{B\in V}C_{AB} \ = \ \sum_{B\in V}\frac{E(B,A)\ + \ I(B,A) }{E(A)\ + \ I(A)}
 \ = \  \frac{E(A)\ + \ I(A) }{E(A)\ + \ I(A)} \ = \ 1.$$

\item A country trading significatively with many different partners will, in general, have low dependence on any particular country.
\end{itemize}

\medskip

\begin{figure}
\begin{center}
\includegraphics[scale=0.4]{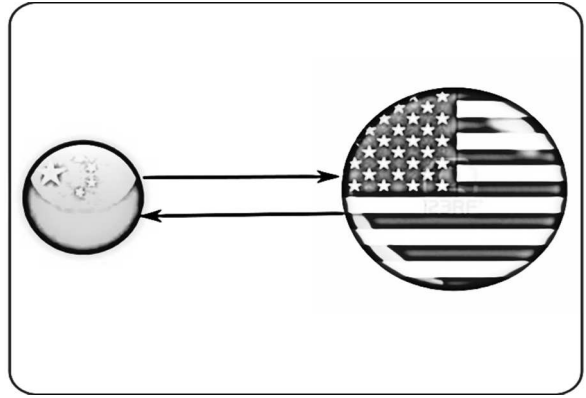}
\caption{China-US bilateral influences.}
\label{hoy}
\end{center}
\end{figure}

US and China are the first and the second world largest economies, respectively. Thus their mutual influences are a topic of global interest, see Figure \ref{hoy} in which the diameter of vertices is roughly proportional to GDP. Definition \ref{ddit} gives us a quantitative method for measuring  these influences.\\

According to US's data its exports and its imports in 2011 (in thousands of US dollars) were 1,479,730,169
$\ \mbox{and} \ $ 2,262,585,634, respectively; while its exports to and its imports from China were 103,878,414 $\ \mbox{and} \ $ 417,302,859, respectively. So the influence of China on the US's international trade is:
$$C(\mbox{US}, \mbox{China})\ = \ \frac{E(\mbox{China},\mbox{US}) \ + \ I(\mbox{China},\mbox{US}) }
{E(\mbox{US})\ + \ I(\mbox{US})} \ = \ 0.139. $$
Thus a $13.9\%$ of the US international trade involves China (either as a buyer or as a seller.)\\

According to  China's data its exports and imports in 2011 were 1,898,388,435 $\ \mbox{and} \ $ 1,743,394,866, respectively; while its exports to and its imports from the US were 325,010,987 $\ \mbox{and} \ $ 123,124,009, respectively. So the influence of the US on China's international trade is:
$$C(\mbox{China}, \mbox{US})\ = \ \frac{E(\mbox{US},\mbox{China}) \ + \ I(\mbox{US},\mbox{China}) }
{E(\mbox{China})\ + \ I(\mbox{China})} \ = \ 0.123. $$
Thus a $12.3\%$ of China's international trade involves the US.\\

So, with respect to trade, China has a higher direct influence on the US than viceversa, and influences
in both directions are quite high.

\section{Direct Influences on Offer}\label{dio}

The definition of direct influences given in the previous section measures international trade.
In this section we take a different approach that considers both international trade (trough exports and imports)
and domestic trade (trough GDP).
\begin{definition}\label{ddio}
The direct influence $D_{AB}$ of country $B$ on the total offer of country $A$ is given by:
$$D_{AB}\ = \ \frac{E(B,A)\ + \ I(B,A) }{\mbox{GDP}(A)\ + \ I(A)}$$ where:
\begin{itemize}
\item$E(B,A)$ is the amount of exports from $B$ to $A$, according to $A$'s data.
\item  $I(B,A)$ is the amount of imports from $B$ to $A$, according to $A$'s data.
\item GDP$(A)$  and $I(A)$ are $A$'s gross domestic product and imports total, respectively.

\end{itemize}
\end{definition}

The direct influence on offer $D_{AB}$ measures the relative size of the trade between $A$ and $B$ in comparison with
$A$'s total offer GDP$(A) + I(A)$. In other words, $D_{AB}$ is the  portion
of products trade in $A$ (both home grown and imported) that involve  $B$
(either because they were imported from or exported to $B$). \\

A couple remarks on the definition of direct influences on offer $D_{AB}$:\\

\begin{itemize}
\item The total dependence $d_A$ of country $A$ measures the portion of its offer that involves
(trough imports or exports) the rest of the world. Indeed we have that
$$d_A \ = \ \sum_{B\in V}D_{AB} \ = \ \sum_{B\in V}\frac{E(B,A)\ + \ I(B,A) }{GDP(A)\ + \ I(A)} \ = \
\frac{E(A)\ + \ I(A) }{GDP(A)\ + \ I(A)} .$$

\item The greater  the GDP of a country, the lesser its dependence on other countries.

\end{itemize}

\medskip

Let us compute the China-US bilateral influences on offer. The GDP of China and US  amounted in 2011 to 7,321,935,025 $\ \mbox{and} \ $ 14,991,300,000, respectively. Thus  the influence of China on the US's offer is given by:
$$D(\mbox{US}, \mbox{China})\ = \ \frac{E(\mbox{China},\mbox{US}) \ + \ I(\mbox{China},\mbox{US}) }
{\mbox{GDP}(\mbox{US})\ + \ I(\mbox{US})} \ = \ 0.0302. $$
Thus a $3.02\%$ of the US's offer involves China (either as a buyer or as a seller.)\\

The influence of the US on China's offer is in turn given by:
$$D(\mbox{China}, \mbox{US})\ = \ \frac{E(\mbox{US},\mbox{China}) \ + \ I(\mbox{US},\mbox{China}) }
{\mbox{GDP}(\mbox{China})\ + \ I(\mbox{China})} \ = \ 0.0494. $$
Thus a $4.94\%$ of China's offer involves the US.\\

So, regarding offer, we find that the US have a higher influence
on China than viceversa, reversing the result obtained for direct influences on trade in Section \ref{dit}.

\section{Indirect Influences on Trade Networks}\label{hiitn}

To account for indirect influences on a trade network we apply the PWP method from Section \ref{mcii} to the
matrices $C$ and $D$ of Definitions \ref{ddit} and \ref{ddio} measuring direct influences on trade and offer.  Thus we obtain the matrices indirect influences on trade and offer
given, respectively, by
$$T(C)\ = \ \frac{e^{\lambda C}_{+}}{e^{\lambda}_{+}}  \ \ \ \ \ \ \ \mbox{and} \ \ \ \ \ \ \  T(D) \ = \ \frac{e^{\lambda D}_{+}}{e^{\lambda}_{+}}. $$

Let us consider in details an extreme case that illustrates the importance of taking indirect influences into account. We use data from the world trade network to be discussed in Section \ref{wtn2}.\\

For political reasons Cuba and the US have fairly low direct trade, and as a result the direct influence of the US on Cuba in trade and offer are just $0.03$ and $0.006$, respectively, i.e. US is directly involved in $3\%$ and $0.6\%$ of the Cuban international trade and offer, respectively. Comparing direct and indirect influences of the US on Cuba one obtains, after normalization, increments of $66\%$ and $20\%$ on trade and offer, respectively. These increments in indirect vs direct influences are quite remarkable since, as we shall see in Section \ref{wtn2}, the US has
decreasing global indirect influences both on trade and on offer. \\

\begin{figure}[h]
\begin{center}
\includegraphics[width=7cm,height=9cm]{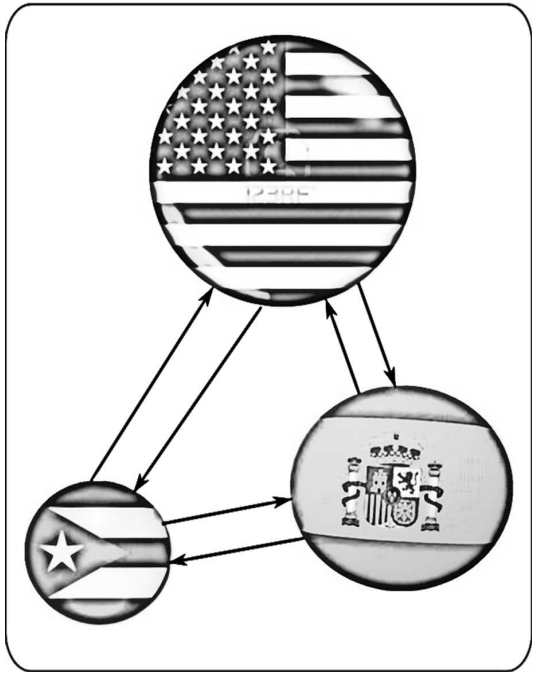}
\caption{US-Spain-Cuba triangle of direct influences.}
\label{e1}
\end{center}
\end{figure}

The increments in indirect influences come from the fact that, beyond direct trade, the US also influences Cuba trough chains of direct trade, which in the simplest case take the form of triangulations.
For example, in  Figure \ref{e1} we show the triangle of direct influences US-Spain-Cuba:  the US exerts a direct influence on Spain in trade and offer of $3.8\%$ and $1.3\%$, respectively;  in turn, Spain exerts a direct influence on Cuba in trade and offer of $6.8\%$ and $1.2\%$, respectively. Thus it is quite clear that the US economy exerts an indirect influence on the Cuban economy trough Spain. \\

Computing indirect influences gives us a method for quantifying chains of direct influences.
In the next sections we compute indirect influences on trade networks for a couple of scenarios: first we consider the American continent with 35 countries, and then we consider the world wide case with 177 countries. In both cases we compute direct and indirect influences, dependencies, and connectedness  starting both from the weight based on trade, and
the weight based on offer. This generates various indexes for gauging influences in international trade. We make a comparative analysis of the different rankings when applied to these scenarios.

\section{American Continent Trade Network}\label{secactn}

We consider 35 countries located on the American continent ranging from mighty US to tiny Dominica.\footnote{ Antigua and Barbuda, Argentina, Bahamas, Barbados, Belize,
Bolivia, Brazil, Canada, Chile, Colombia, Costa Rica, Cuba, Dominica, Dominican Republic, Ecuador, El Salvador, United States, Grenada, Guatemala, Guyana, Haiti, Jamaica, Mexico, Nicaragua, Panama, Paraguay, Peru, Saint Kitts and Nevis, Saint Lucia, Saint Vincent and the Grenadines, Suriname, Trinidad and Tobago, Uruguay, and Venezuela.}
The trade network for the American continent is a nearly complete graph with
each country trading in average with 32.2 countries out of 34 possibilities. The fully connected countries are Argentina, Brazil, Canada, Colombia, Costa Rica, Cuba, Dominican Republic, Ecuador, Guatemala, Jamaica, Mexico, Peru, Salvador, and the US. The lesser connected countries are  Bolivia, Dominica, Haiti, and Paraguay,   trading respectively with $29, 30, 12,$ and $27$ countries.\\

\begin{figure}[h]
\begin{center}
\includegraphics[width=12cm,height=13cm]{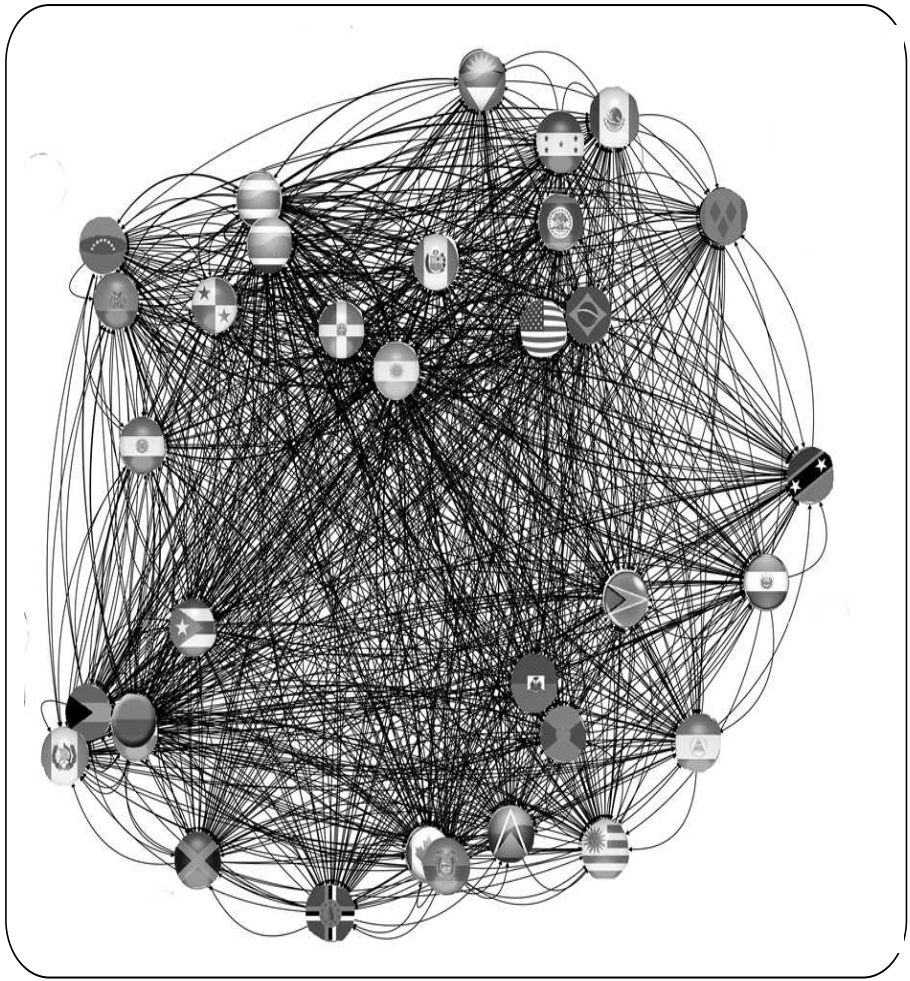}
\caption{American Continent Trade Network: countries displayed by their flags.}
\label{casa}
\end{center}
\end{figure}

The American continent trade network, shown  in Figure \ref{casa},  was drawn with SAGE  using the matrix $D_{AB}$ of direct influences on offer  from Definition \ref{ddio}.\\

\begin{figure}[h]\label{atnit}
\begin{center}
\includegraphics[width=12cm,height=7cm]{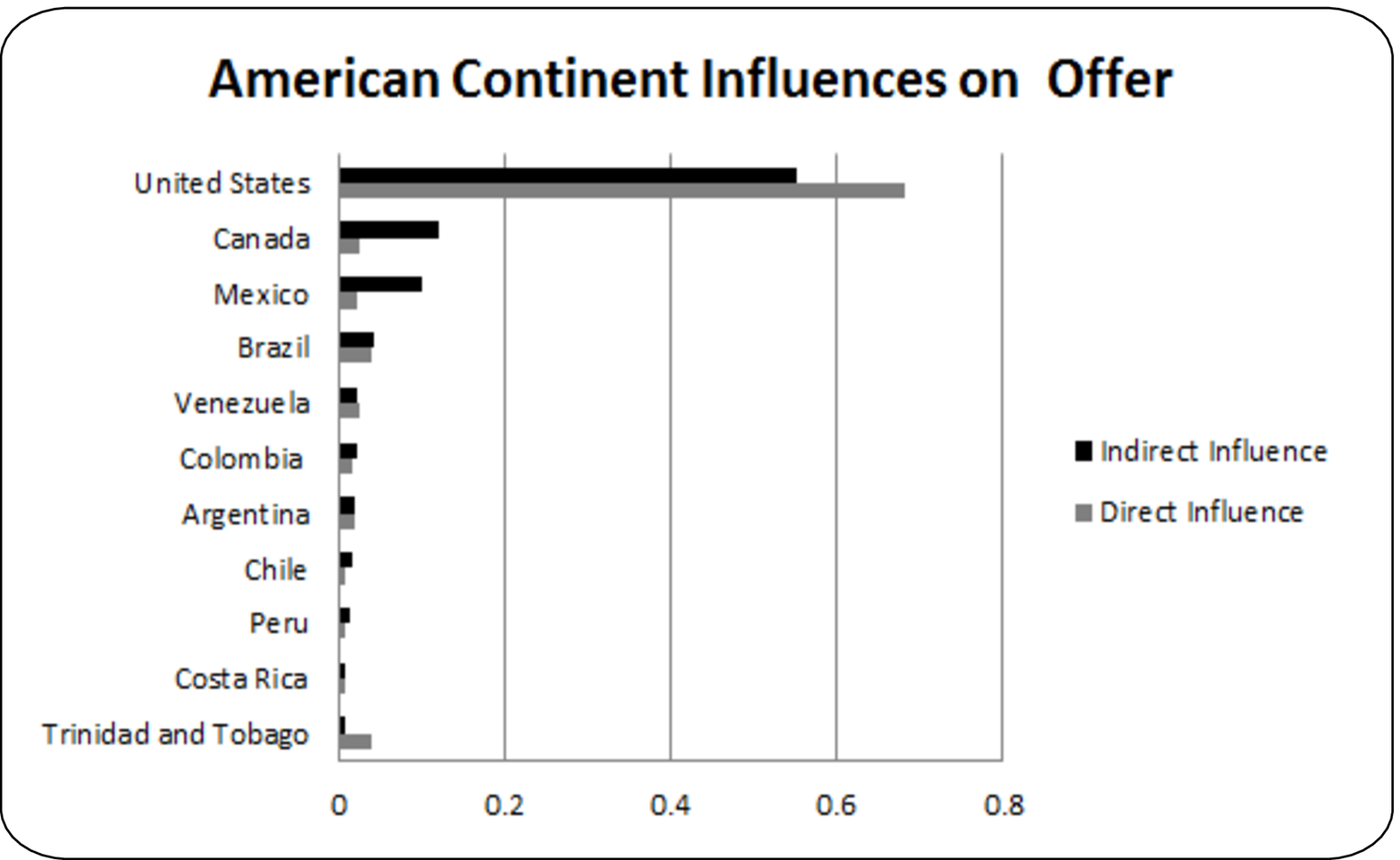}
\caption{Influences on Offer in the American Continent Trade Network}
\label{g2}
\end{center}
\end{figure}

The first thing one notices from the table in Figure \ref{g2} is that the US, as expected, leads both in direct and indirect influences. One, however, may wonder why its direct influences are higher than its indirect influences? This happens, we believe, for a couple of reasons: 1) Because of its leading role, the US is more likely to trade with countries otherwise isolated with
little influence on other countries. 2) Since influences are normalized, the decreasing indirect influences of the US may also be explained by the opposite behaviour in countries such as Canada and Mexico, whose indirect influences increase to a good extend precisely because of their special relation (NAFTA agreement) with the leading country.\\

Note that Brazil has higher direct influences than both Canada and Mexico, this is to be expected since Brazil's GDP is only second to the US's in the American continent. Nevertheless, Brazil has lesser indirect influences than both Canada and Mexico. This is due to the fact that in comparison Brazil is an inward looking economy, and does not have a privileged relation with the leading country.\\

The case of Chile is interesting as well. Its direct influences are quite modest, since Chile has a relatively small economy. However, it has a noticeable increment in indirect influences,  due to the fact that Chile has an export-oriented economy.\\

Trinidad and Tobago (TT) is perhaps the most revealing case. Its small size economy seems hardly compatible with its high direct influence, and to make this case even more intriguing its indirect influences decrease quite sharply. This curious situation may be explained as follows. Our indices measure influences on countries, and TT exerts a direct influence on quite a few low GDP countries,  mostly Caribbean Community (CARICOM) member states. Since CARICOM's countries have small economies the influence of TT on them may be quite high, even if TT has itself a relatively small economy. On the other hand, many CARICOM member states have low direct influences, so they seldom contribute to the indirect influences of TT. In summary, the difference between direct and indirect influences of TT  attests to both the leadership of TT within CARICOM, and the meager impact of the subregion in the economy of the American continent at large.\\

\begin{figure}[h]\label{actn}
\begin{center}
\includegraphics[width=12cm,height=7cm]{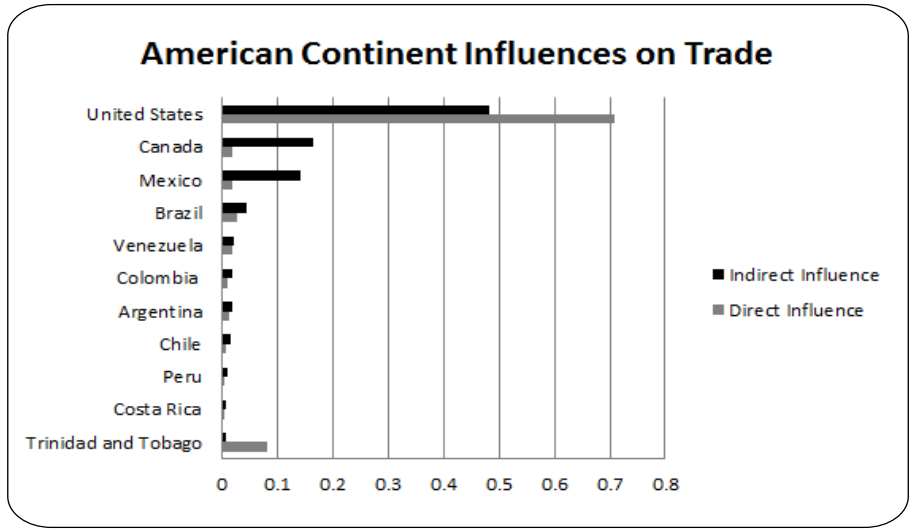}
\caption{Influences on Trade in the American Continent Trade Network}
\label{g1}
\end{center}

\end{figure}

Considering influences on trade in the American continent, see Figure \ref{g1},   we find even less correlation between direct and indirect influences. Canada and Mexico indirect influences increase quite sharply, while for  TT there is a deep fall instead. Antigua goes
from position $11$ in direct influences, to position $31$ in indirect influences. Bolivia has the opposite behaviour going from position $32$ in direct influences to position $20$ in indirect influences.

\section{World Trade Network}\label{wtn2}

\begin{figure}[h]
\begin{center}
\includegraphics[width=12cm,height=15cm]{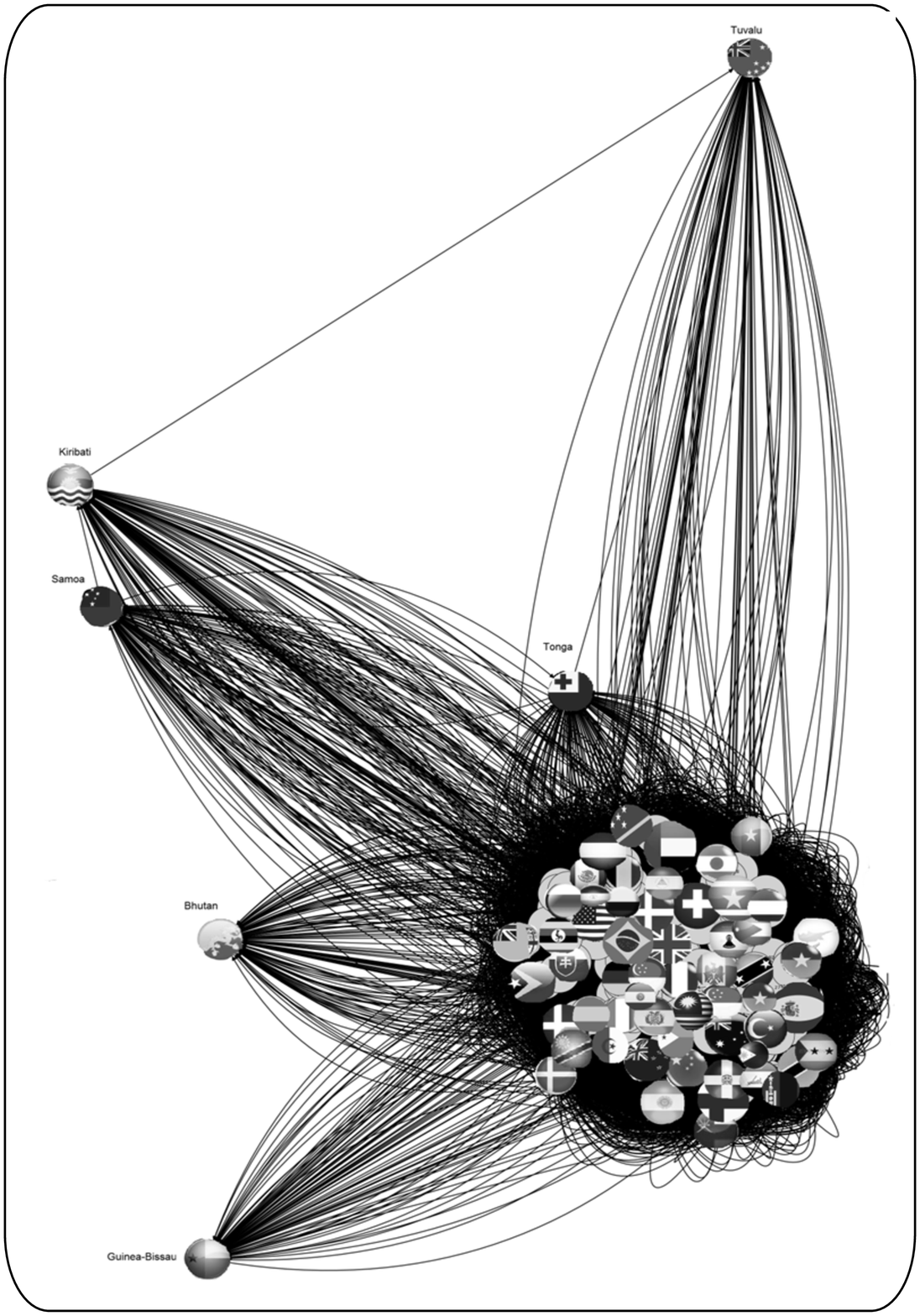}
\caption{World Trade Network: most countries take part in a giant connected component, with the exception of a few isolated countries such as Buthan, Guinea Bissau, Kiribati, Samoa, Tonga, and Tuvalu.}
\label{casawtn}
\end{center}
\end{figure}

The world trade network containing 177 countries distributed all around the globe, shown schematically  in Figure \ref{casawtn}, was drawn feeding SAGE with the matrix of direct influences on offer from Definition \ref{ddio}. \\

Unsurprisingly, the world trade network contains a big highly connected component containing the vast majority of countries in the
world, with the exception of a handful of commercially isolated countries such as Buthan, Guinea Bissau, Kiribati, Samoa, Tonga, and Tuvalu. On average a country trades with $131.46$ countries, and there are $24$ countries each trading with at least $173$ countries. \\

We proceed to address a couple of questions concerning the world trade network: Are GDP and direct or indirect influences correlated? Are there significative differences between direct and influences? These question are both analyzed by looking first to influences on offer, and then influences on trade. Instead of a statistical analysis we simply highlight interesting cases regarding the top 20 countries in indirect influences.  \\

\begin{figure}[h]
\begin{center}
\includegraphics[width=12cm,height=10cm]{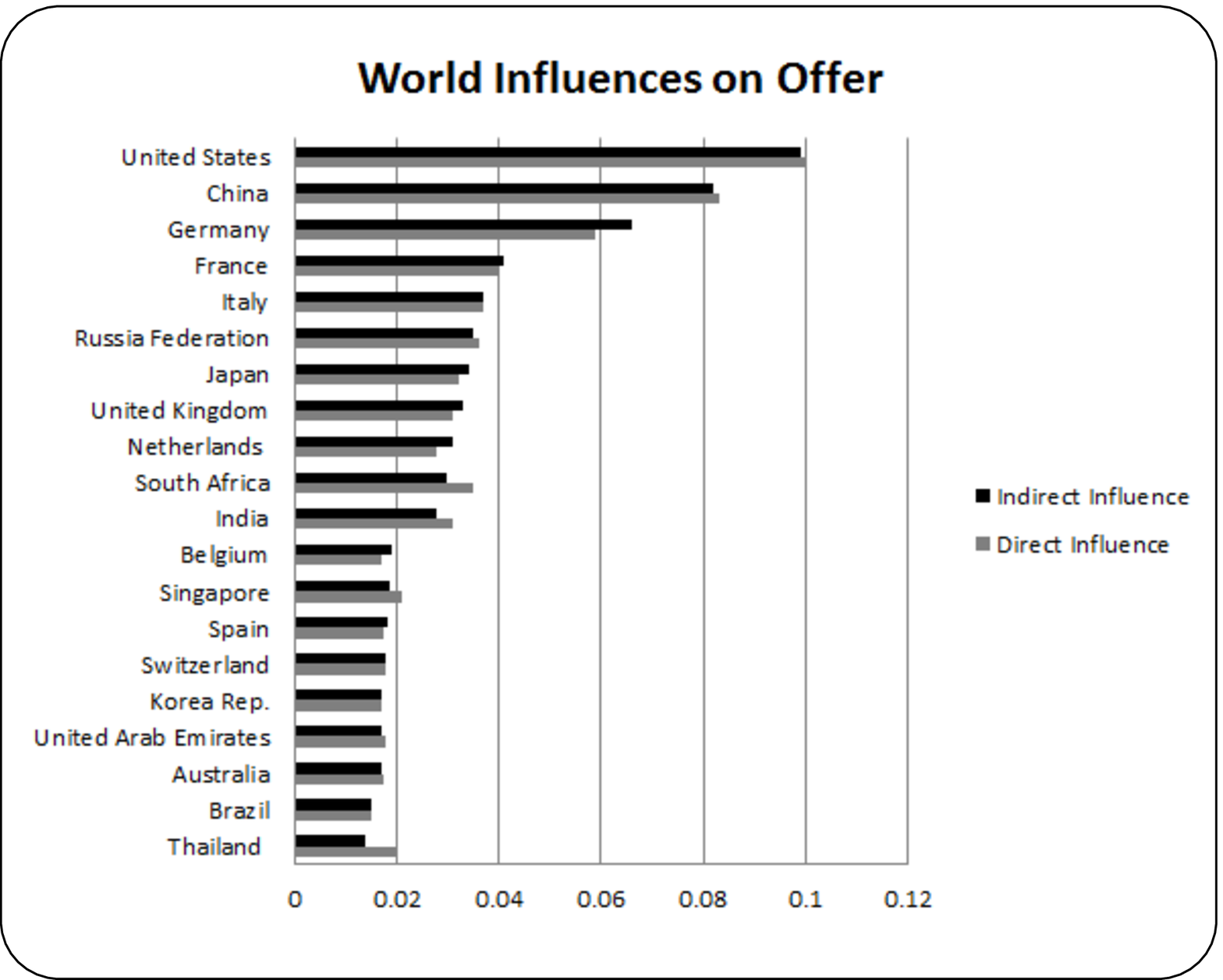}
\caption{Influences on Offer in the World Trade Network.}
\label{casag2}
\end{center}
\end{figure}

Naturally, we expect high GDP countries to have high direct and indirect influences on offer, and indeed there is a generic tendency for this correlation to hold, however the actual data reveals a far more subtle situation, see Figure \ref{casag2}.\\

Japan has the $3$rd largest world GDP, but occupies the $8$th and $7$th places in direct and indirect influences, respectively. Canada is in  the $11$th place in the GDP ranking, but occupies the 20th and 21st positions in the direct and indirect influences rankings, respectively. Similarly, Mexico is in the 14th place in the GDP ranking, but 34th and 31st positions in direct and indirect influences. The drop in influences compared to GDP in the latter cases is probably  consequence of Canada's and Mexico's strong economic ties to the US, so their influences are somewhat diluted in the US high GDP.\\

Comparing direct and indirect influences on offer, see Figure \ref{casag2}, we find countries with similar results in both rankings, for example
Brazil, Italy, Switzerland, and Korea. There are countries with diminishing indirect influences as Australia, China, India, Russia, Singapore, South Africa, Thailand, United Arab Emirates and the US. In the case of South Africa and even more so for Thailand the decreases are quite sharp. There are also countries with raising indirect influences as Germany, France, Japan, United Kingdom, Netherlands, Belgium and Spain. As a rule European countries tend to have increasing indirect influences, the case of Germany being remarkable because a sharp increase occurs. \\

\begin{figure}[h]
\begin{center}
\includegraphics[width=12cm,height=10cm]{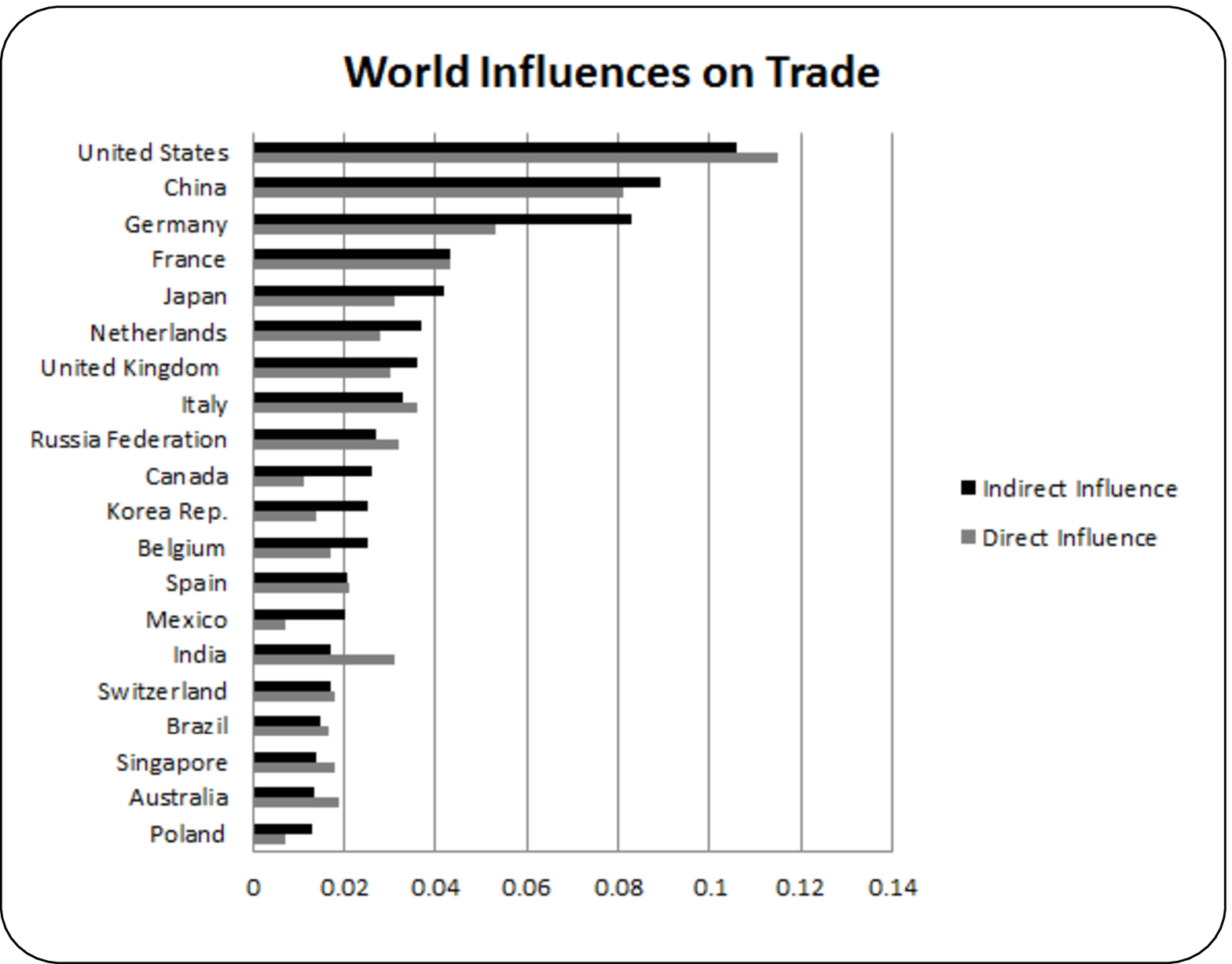}
\caption{Indirect Influences on Trade in the World Trade Network}
\label{casagra}
\end{center}
\end{figure}

Next we consider direct and indirect influences on trade in the world trade network, see Figure \ref{casagra}. In this case we expect less correlation with the GDP ranking, since we are looking at influences on international trade and disregarding domestic trade. For example, Netherlands climbs from position $17$ in GDP to position $7$ in indirect influences on trade. Similarly, Belgium occupies the $24$th position in GDP, but raises to the $12$th position
in indirect influences on trade. In the opposite trend, Brazil is in the $6$th place in GDP, but is in the $17$th position in indirect influences on trade.\\

From  Figures \ref{casag2} and \ref{casagra}, it is quite clear that influences on trade measure a different thing that influences on offer, even if the US, China, Germany and France, in this order, remain the top 4 countries in both rankings. For example, South Africa, Thailand, and the United Arab Emirates are not longer in the top 20 countries in influences on trade, and are replaced by Canada, Mexico and Poland. Observe that Brazil, Japan, Korea, Netherlands, Spain, and the United Kingdom  climb in their ranking positions, Belgium remains stable, and Australia, India, Italy, Russia, Singapore, and Switzerland fall in their rankings positions.\\

Similarly, direct and indirect influences on trade are weaklier correlated than direct and indirect influences on trade.
Only France, among the top 20 countries, have nearly identical direct and indirect influences on trade.
Australia, Brazil, India, Italy, Russia, Singapore, Spain, Switzerland, and the US have diminishing indirect influences, whereas
Belgium, Canada, China, Germany, Japan, Korea, Mexico, Netherlands, and the United Kingdom have increasing indirect influences.
The cases of Germany and Korea are interesting because of their sharp increments.\\

Below we show the indirect dependence-influence planes, both for trade and offer, of the world trade network. Note that the United States and China have high influence and low dependence on offer, while they  have both high influence and dependence on trade, this is due to their high GDPs. In contrast, Germany have high influence and high dependence both on offer and on trade.

\begin{figure}[h]
\begin{center}
\includegraphics[width=12cm,height=8cm]{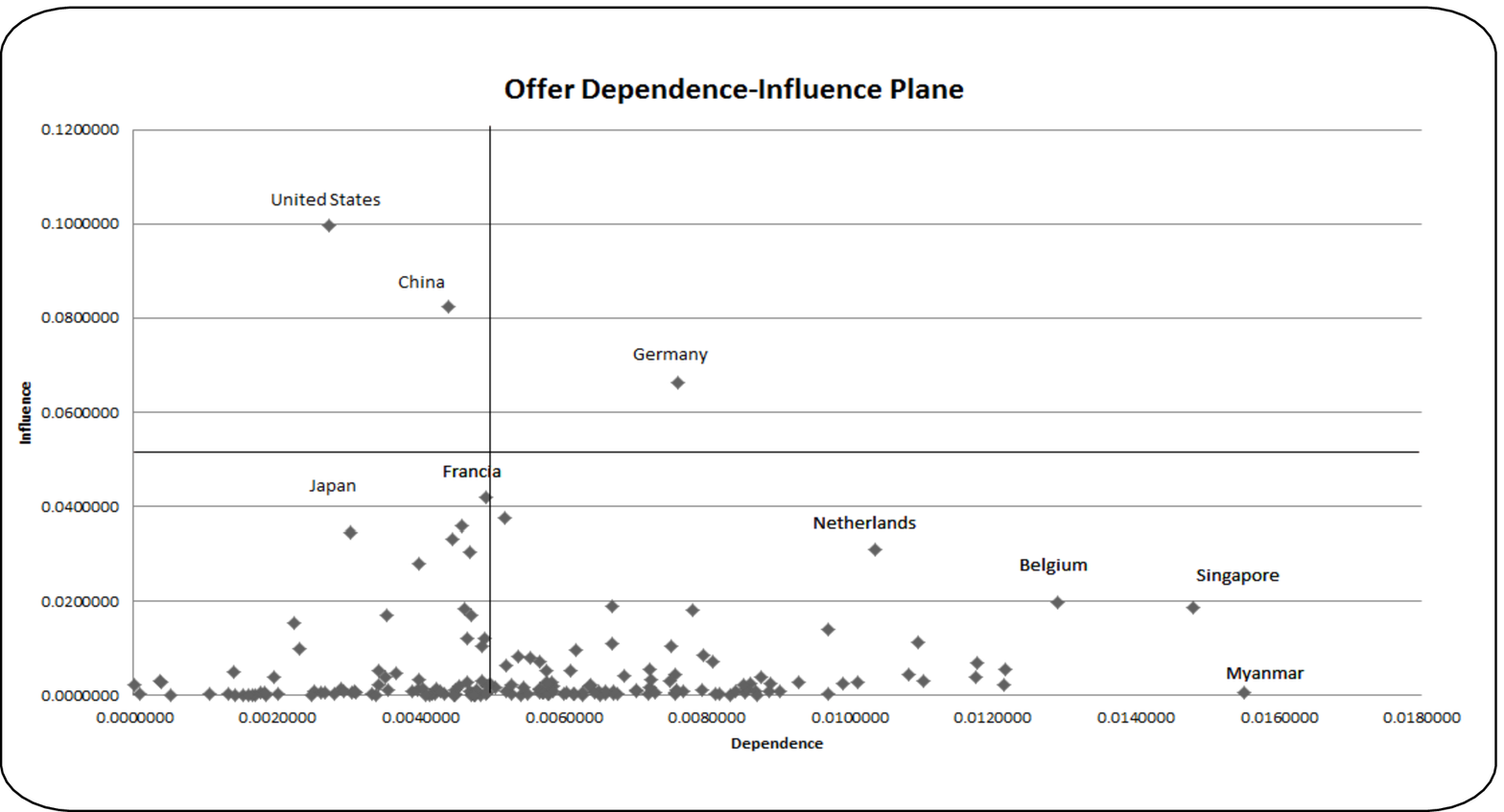}
\caption{Offer dependence-influence plane}
\label{gr1}
\end{center}
\end{figure}

\section{Conclusion}

In this work we developed a network viewpoint leading to a couple of ways of ranking direct influences of countries in the international trade market, namely, a ranking based on trade and another ranking based on offer. The first ranking measures relative international trade trough exports and imports, while the second ranking also takes into account domestic trade trough GDPs.\\

As far as we can tell a network approach to international trade taking into account indirect influences has not been designed so far. Most studies found in the literature gauge direct influences as evidenced by trade accounts. In this regard, they compare to our networks of direct influences in trade, rather than to our networks of indirect influences in trade.
We showed that taking in consideration indirect influences may lead to more realistic indexes, as relevant cases suggest.\\

We computed indirect influences using the PWP method, and applied it to the American continent trade network and to the World trade network. Thus we obtained new rankings and the corresponding dependence-influence planes. \\

\begin{figure}
\begin{center}
\includegraphics[width=12cm,height=8cm]{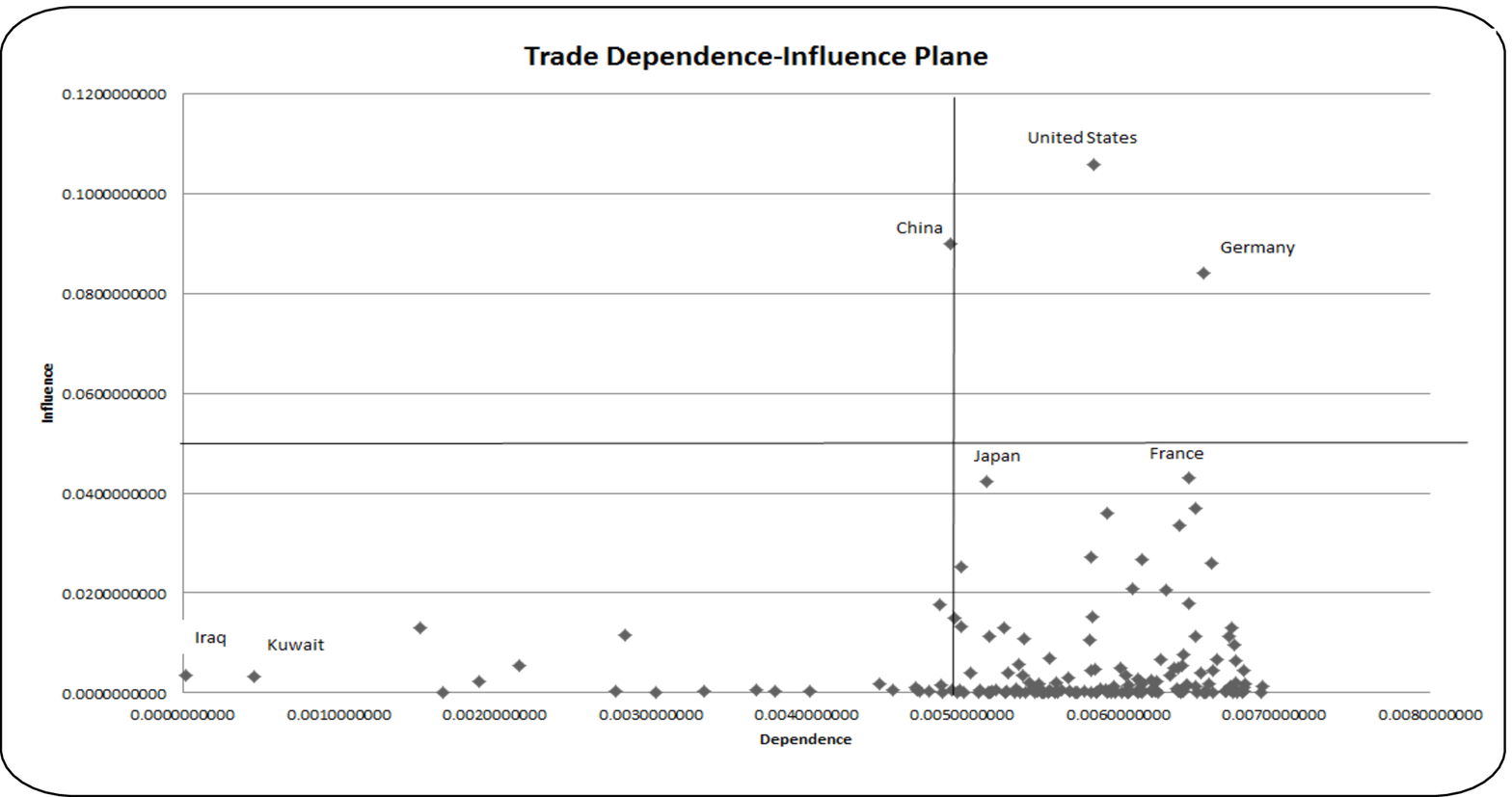}
\caption{Trade dependence-influence plane}
\label{gr2}
\end{center}
\end{figure}

Part of the motivation for this work is the high public interest in ranking countries for their centrality in international trade. Various rankings have been introduced, some quite sophisticated such as the DHL Global Connectedness Index  \cite{r12}. This ranking however results from the integration of many features, whereas our rankings focus on a single feature enhanced by taking into account indirect influences. \\

The problem of further integrating the various rankings proposed in the literature may be attacked by thinking of the set of
rankings as a finite metric space.  For example, we can define a suitable normalized Euclidean distance between indices $r,l : V \longrightarrow [ 1, n ] $ (with $r(A)$ and $l(A)$ being the ranking positions of country $A$ according to $r$ and $l$, respectively; and $n$ is the total number of countries) given by
$$ d(r,l) \ = \  \frac{1}{( n - 1) \sqrt{ n}}\sqrt{ \sum_{A \in V } ( r(A) - l(A))^{2}} .$$
Furthermore, we also have the problem of consolidating the various indexes into just one index.\\

We applied our methods for computing indirect influences to trade networks at the level of countries. It should however be clear that these methods may also be applied at the business level, and even at the individual level.
In the latter cases, finding reliable data at a global scale and managing
such huge data are daunting problems. Nevertheless, our methods can be readily applied if one focusses on specific
business sectors, just like we made a focus study of indirect influences on trade and offer in the American continent.\\

The  models presented in this work were static, in the sense that our data referred to just one year, namely 2011. Nevertheless, our techniques may be extended to a multiple years  dynamical model in which influences are time depended functions.

\end{document}